\documentclass[twocolumn,showpacs,preprintnumbers, floatfix]{revtex4}
\usepackage{amsmath}
\usepackage{graphicx}

\begin{document}

\title{Organization of atomic bond tensions in model glasses}
\author{T. Kustanovich and Y. Rabin}
\affiliation{Department of Physics, Bar Ilan University, Ramat-Gan 52900}
\author{Z. Olami}
\affiliation{Department of~~Chemical Physics, The Weizmann Institute of Science, Rehovot 76100, Israel}

\begin{abstract}
In order to understand whether internal stresses in glasses are correlated
or randomly distributed, we study the organization of atomic bond tensions
(normal forces between pairs of atoms). Measurements of the invariants of
the atomic bond tension tensor in simulated 2D and 3D binary Lennard--Jones
glasses, reveal new and unexpected correlations and provide support for
Alexander's conjecture about the non-random character of internal stresses
in amorphous solids.
\end{abstract}
\pacs{61.43.-j 61.43.Fs    }

\maketitle

\section{Introduction}

Although the existence of internal stresses in glasses is well-established %
\cite{Egami}, little is known about their statistical properties and about
their role in the physics of glasses and amorphous solids. Intuitively, one
expects that since local stresses and strains in a glass are a by-product of
the structural disorder produced by rapid freezing of a liquid (or of
incompatible local packing and space-filing requirements \cite{Kleman} \cite{Kivelson}),
their distribution should be random. A different view is advocated in ref. %
\cite{Alexander}, where it is argued that the local stresses are not random,
and that their organization is closely related to the low temperature
vibrational spectra of glasses. Even though the question whether internal
stresses in glasses are random\cite{Egami97} or are subject to some complex
ordering is of fundamental importance, to the best of our knowledge this
issue was not directly addressed by experiments to date. A possible reason
is that the effects of internal stresses are discernible only on atomic
length scales, whereas on larger, but still microscopic length scales, they
average to zero and the usual description of a glass as an isotropic solid
with zero average internal stress, applies.

Unlike the stress singularities produced by lattice defects (vacancies,
dislocations) in crystalline solids, internal stresses in glasses cannot be
attributed to strongly strained core regions embedded in a weakly strained
bulk. As a consequence, classical elastic theory methods for the calculation
of strains and of the resultant stresses\cite{LL}, are not applicable to
glassy systems. The problem has been solved for periodic solids subjected to
arbitrary homogeneous stress, for which one can extend the continuum
approach down to atomic scales and calculate the local stress acting on a
primitive cell\cite{Born}. Lately, it was shown that if the internal
stresses in an amorphous solid vary sufficiently slowly on microscopic
length scales, one can obtain a similar (to the crystalline case) expression
for the local internal stress, with the volume of the primitive cell
replaced by the Voronoi volume\cite{Alexander}. However, recent numerical
studies of atomic stresses in two and three dimensional glassy systems at
mechanical equilibrium, have demonstrated that the atomic environments vary
rapidly even between neighboring atoms at zero temperature \cite%
{Kustanovich1,Wittmer,thesis}. The breakdown of the continuum approach
implies that the macroscopic concept of stress can not be extrapolated down
to atomic scales in glassy solids.

In order to bypass these difficulties, in this paper we introduce the atomic
bond tension tensor that accounts for the magnitudes and the directions of
the normal forces (bond tensions in the terminology of ref. \cite{Alexander}%
) acting between pairs of atoms with radial pair interaction. Since this
tensor can not be identified with the stress tensor, the original question
about the random vs. non-random character of the internal stresses, is
replaced by that concerning the organization of atomic bond tensions in
glasses.

In section II we define the atomic bond tension tensor and discuss its
physical meaning. In section III we describe the measurement of this
parameter in simulated two and three dimensional binary Lennard-Jones
Inherent States of liquids (using the terminology of ref. \cite{Waber}) and
of glasses in mechanical equilibrium. In section IV we show that the local
organization of bond tensions belonging to an atom is not random and that
the observed correlations between the eigenvalues of the proposed tensor
have a physical meaning. In section V, directional correlations and
clustering are addressed, and some parallels with granular matter are drawn.
These results, which demonstrate that the atomic bond tension tensor probes
simulated glasses in a new way, thus extracting information unavailable by
the existing methods, are discussed in section VI.

\section{The atomic bond tension tensor}

Consider a classical system of atoms with coordinates $\{\overrightarrow{R}%
_{i}\},$ interacting through the pair potential $U(\left\{ R_{ij}\right\} )$%
, where $R_{ij}=|\overrightarrow{R}_{i}-\overrightarrow{R}_{j}|$ is the
distance between atoms $i$ and $j$. The tension in the ``bond'' $[ij]$
between the $i-$th and the $j-$th atom is defined as 
\begin{equation}
J_{ij}=\frac{dU(R_{ij})}{dR_{ij}}.  \label{Jij}
\end{equation}%
Notice that in a perfect monatomic crystal, all bond tensions vanish in
mechanical equilibrium, whereas in a multi--component crystal there are $%
J_{ij}\neq 0$. Non-vanishing $J_{ij}$ are indications of ``mismatches'' in
the bonds.

Whereas the definition of bond tensions (or, equivalently, of the normal
forces) is strictly local, their distribution, $P(J_{ij})$, depends on the
spatial order. $P(J_{ij})$ is related to the radial distribution function $%
g(r)$ by $P(J_{ij})dJ_{ij}=g(r)dr$, where $P(J_{ij})dJ_{ij}$ is the
probability to find bond tension with values between $J_{ij}$ and $%
J_{ij}+dJ_{ij}$, and $g(r)dr$ is the probability of finding a particle
between $r$ and $r+dr,$ given a particle at the origin. Therefore, the
variation in nearest neighbor separations, manifested in the width of first
peak of $g(r)$, leads to a broad distribution of bond tensions (the 
distribution of the normal forces was calculated for granular materials
subjected to external forces\cite{Liu}, \cite{Radjai}, and for liquid and
glassy systems\cite{OHern}). The above connection between the
distributions $P(J_{ij})$ and $g(r)$ has several important implications.
If the separations $R_{ij}$ between an atom and its neighbors are
randomly chosen from the distribution $g(r)$, one does not expect to find
correlations between the corresponding $J_{ij}$. If separations between
nearest-neighbors are nearly equal, the corresponding bond tensions must be
nearly equal as well (the atomic environments are either isotropically
stretched or isotropically compressed).

Although there are numerous possible realizations of the set of interatomic
distances $\{R_{ij}\}$, their number is limited by several constraints. In
mechanical equilibrium, the atomic positions are constrained by the
requirement that the net force on every atom in the system vanishes, i.e., 
\begin{equation}
\overrightarrow{f}_{i}=\sum_{j}J_{ij}\hat{R}_{ij}=0,  \label{force}
\end{equation}%
where $\hat{R}_{ij}$ is a unit vector in the direction of $\overrightarrow{R}%
_{ij}$. Moreover, mechanical stability requires that every atom in the
system is located at its local energy minimum. That means that not only the
Einstein frequencies of every atom in the system but also all the
eigenvalues of the complete dynamic matrix are positive \cite{Alexander}.
These requirements must be satisfied by Inherent States of a liquid (in the
terminology of ref. \cite{Waber}) as well as by glasses at zero temperature.
In the case of granular matter thermal fluctuations play no role and only
the requirement of force balance applies at rest \cite{Edwards}, \cite%
{Coppersmith}. This is an important difference, especially in view of a
suggestion due to Alexander \cite{Alexander}, that the structural disorder
in glasses at zero temperature should be constrained by the stability
requirements.

If the distribution of bond tensions belonging to an atom $i$ is not
isotropic, the directions associated with the different bonds should be
taken into consideration. This is done by introducing the tensor $T_{ij}$
(with $i\not=j$) with components 
\begin{equation}
T_{ij}^{\alpha \beta }=J_{ij}\;\hat{R}_{ij}^{\alpha }\;\hat{R}_{ij}^{\beta }.
\end{equation}%
In order to describe the bond tensions acting on an atomic environment, the
site of atom $i$ is taken as a reference point and summation over all $%
T_{ij} $ (with $j\neq i$) is performed. The resulting \emph{atomic bond
tension tensor} $\overleftrightarrow{T_{i}}$ is given by 
\begin{equation}
\overleftrightarrow{T_{i}}=\sum_{j}\overleftrightarrow{T_{ij}}.\;_{.}
\label{tai}
\end{equation}%
The physical meaning of $\overleftrightarrow{T_{i}}$ become apparent when it
is decomposed to its trace 
\begin{equation}
J_{i}=\sum_{j}J_{ij}  \label{J}
\end{equation}%
and to the traceless part $\overleftrightarrow{L_{i}}$ , with components 
\begin{equation}
L_{i}^{\alpha \beta }=\sum_{j}J_{ij}\;\left( \hat{R}_{ij}^{\alpha }\;\hat{R}%
_{ij}^{\beta }-\frac{1}{d}\delta _{\alpha \beta }\right)  \label{L}
\end{equation}%
where $d$ is the spatial dimensionality of the system and $\delta_{\alpha \beta}$ is the
Kronecker delta function. If the distribution
of atomic bond tensions (ABT) is isotropic, the bond tensions acting on an
atom $i$ are completely described by the trace $J_{i}$ that can be
interpreted as the effective pressure acting on the atomic environment \cite%
{Kustanovich2}, \cite{thesis}. The traceless tensor $\overleftrightarrow{%
L_{i}}$, Eq. \ref{L}, describes the effect of anisotropy on the ABT tensor.
To illustrate this point, consider a monatomic crystal. At mechanical
equilibrium with no external stresses, all bond tensions vanish. If the
system is under hydrostatic compression, the atomic environments vary
isotropically so that $J_{i}\not=0,$ while all the components $L_{i}^{\alpha
\beta }$ vanish. Under pure shear, $J_{i}$ vanishes whereas $L_{i}^{\alpha
\beta }\not=0$ and we conclude that $\overleftrightarrow{L_{i}}$ describes
the local effect of shear. Since $\overleftrightarrow{L_{i}}$ and $J_{i}$
are related to local shear and local pressure, respectively, for monatomic
crystals $T_{i}$ can be identified (upon suitable normalization by the
surface area of the unit cell) with the local stress. For structurally
disordered systems, the variation in system's volume would lead, in general,
to variations in the local $\overleftrightarrow{L_{i}}$ . If the system is
not sheared as a whole, $\overleftrightarrow{L_{i}}$ measures the weighted
effect of the intrinsic anisotropy, on the distribution of ABT.

Although the ABT tensor, Eq. \ref{tai}, was constructed with the aim to
characterize the bond tensions acting on an atomic environment, its
components are not rotation-translation invariant and, unlike bond tensions,
depend on the choice of the coordinate system. In order to obtain a
basis--independent (scalar) characterization of bond tensions, one can
consider the invariants of the ABT tensor (e.g., in 2 dimensions, the trace
and the determinant). A drawback of such an approach is that among the $d$
invariants ($d$ is the dimensionality of space), only the trace, eq. \ref{J}%
, has the units of bond tension. An alternative approach utilizes the fact
that the eigenvalues of the real symmetric matrix $\overleftrightarrow{T_{i}}
$ are invariant under arbitrary translation and rotation and, therefore, for
every atom $i$ in the system, there is a set of $d$ eigenvalues $\left\{
t_{i}\right\} =\left( t_{i}^{1},\;\cdots \;,t_{i}^{d}\right) $ and $d$
eigenvectors $\left\{ \hat{e}_{i}\right\} =\left( \hat{e}_{i}^{1},\;\cdots
,\;\hat{e}_{i}^{d}\right) $ of the ABT tensor. We stress that although $%
\left\{ t_{i}\right\} $ have the same units as the components of the force
acting on the atom, there is a profound difference between the two; the
eigenvalues of ABT tensor are uniquely defined scalars that can be compared
for different atoms at different times and external conditions (temperature,
pressure). They do not vanish when the force acting on the atom vanishes,
and therefore are suitable parameters to characterize atomic environments in
amorphous systems and in Inherent States of liquids.

The object of the present study is to characterize the organization of
atomic bond tensions in glassy systems. To do so using the ABT tensor, the
distribution of the sets $\left\{ t_{i}\right\} $ should be known for every
atom in the system. Such a distribution is trivial in monatomic crystals
under hydrostatic compression, for\ which all the eigenvalues of ABT tensor
are degenerate. While in multi--component crystalline solids the eigenvalues
are, in general, non-degenerate, the fact that the atoms are arranged on a
periodic lattice implies the existence of correlations between the
eigenvalues. Whether the eigenvalues of the ABT tensor in structurally
disordered materials are degenerate, correlated or randomly distributed, is
an open question. In the following we address this question for two and
three dimensional binary Lennard--Jones glasses in mechanical equilibrium.

\section{The simulated systems}

The simplest (and perhaps the only) way to calculate atomic bond tensions is
by numerical simulations. Since the ABT tensor is a hitherto unexplored
quantity, there are many factors to be considered when choosing the
parameters of the simulated system. Local properties, such as the existence
of short range structural order, or global factors, such as the
dimensionality of the system\cite{Perera}, may affect the generality of the
results. In order to suppress short-range-order effects, we use binary
mixtures of two types of atoms interacting through Lennard-Jones potentials
(results for monatomic systems with a pair potential that leads to local
icosahedral ordering were reported elsewhere\cite{thesis}). Since 2D models
of glasses offer some advantages (in terms of computational load and
visualization options) over the more realistic 3D systems, we consider both
two and three dimensional systems with similar repulsive-attractive pair
potentials.

For 3D glasses we use binary 6-12 Lennard-Jones $80:20$ mixtures of atoms of types
$A$ and $B$ (denoted 3DbLJ) with interaction parameters\cite{Kob}: $\varepsilon^{AA}=1.0,\quad
\varepsilon^{AB}=1.5,\quad\varepsilon^{BB}=0.5,\quad\sigma^{AA}=1.0,\quad
\sigma^{AB}=0.8,\quad\sigma^{BB}=0.88.$ Particles of type $A$ are majority
particles (i.e., $80\%$). The interaction vanishes for $r>2.5\sigma^{AA}$.

For 2D glasses we use binary 6-12 Lennard-Jones $50:50$ mixtures (denoted
2DbLJ) where\cite{OHern}\newline
$\sigma^{AB}=0.5(\sigma^{A}+\sigma^{B})$ for $A,\;B=1,2$ with $%
r\leq4.5\sigma^{AB}$. \newline
All measurements are given in reduced LJ units (for Argon these units
correspond to $\varepsilon=120k_{B}$ and $\sigma=3.4$ \AA). The bond
tensions are given in units of $\varepsilon ^{AA}/\sigma^{AA}$ for the 3DbLJ
systems, and $\varepsilon/\sigma^{1}$for the 2DbLJ systems. All atoms are
assumed to have unit mass.

Liquids and glasses were prepared using standard molecular dynamics
simulations preformed at a constant volume\cite{Allen}. Periodic boundary
conditions were assumed. For each system type we considered two densities: $%
\rho =1.2$ and $\rho=1.4$ for the 3DbLJ systems, and $\rho =0.596$ and $\rho
=0.735$ for the 2DbLJ systems. Such difference in densities corresponds to $%
5\%$ ( $10\%$) difference in the average nearest neighbor separation for the
3DbLJ (2DbLJ) system. The number of atoms $N$ for each 3DbLJ system is $%
16384 $. For 3DbLJ system with $\rho =1.2,$ samples with 4000 atoms were
also considered. Comparison between ABT tensors calculated for the small and
the large systems shows that similar results are obtained in both cases.
 For 2DbLJ systems, we used samples
with $N=4096$ atoms.

Since the binary Lennard--Jones systems used in this study do not tend to
crystallize \cite{Kob},\cite{OHern}, samples at zero temperature can be
prepared either by steepest descent quench from the liquid state (Inherent
States of the liquid), or by obtaining a well--equilibrated glass at a
sufficiently low temperature (at which all atoms are effectively trapped in
their cages), and then using steepest descent quench to yield glassy systems
in mechanical equilibrium. For dense 2DbLJ systems, an additional route of
preparation was used: taking samples of low-density 2DbLJ glasses (at zero
temperature), we decreased the length of the simulation box and rescaled all the
atomic positions by $10\%$. Performing a steepest descent quench on the
rescaled samples (in order to relax the inter-atomic forces) yields dense
2DbLJ systems at mechanical equilibrium. Although the above preparation
routes are not equivalent, we did not find statistical differences (through
comparison of means, ranges and variances) between samples prepared by
different methods. Also, directional correlations related to ABT tensor are
similar, regardless of the preparation method. Details about the preparation
and the distribution functions $P(\left\{ t_{i}\right\} )$ are summarized
elsewhere \cite{thesis}.

\section{Local organization of eigenvalues of atomic bond tension tensor}

Since we are not aware of theoretical models that predict correlations
between bond tensions in structurally disordered systems, we first address
the question: Is the organization of the $d$ eigenvalues of the ABT tensor
of an atom random? If the atomic environment is such that it can be
reproduced by a random choice of $d\,\,$numbers from the complete
distribution function $P(\left\{ t_{i}\right\} )$, then the directional
information contained in the ABT tensor is redundant. Observation of
non-random organization, on the other hand, would confirm that the bond
tensions acting on an atomic environment are correlated.

First we show that eigenvalues belonging to the same atom are not degenerate
at zero temperature. This is an interesting point because the non-degeneracy
of the $d$ eigenvalues of ABT tensor implies that the corresponding $d$
eigenvectors form an orthogonal basis, i.e. $\hat{e}_{i}^{\alpha }\cdot 
\hat{e}_{i}^{\beta }=\delta _{\alpha \beta }$. Since in structurally
disordered systems the atomic bond tensions corresponding to any two (even
nearby) atoms $i$ and $j$, may be very different, one does not know a priori
how to relate between the corresponding local coordinate systems $\{\hat{e}%
_{i}\}\,$and $\{\hat{e}_{j}\}$. If, however, the eigenvalues of ABT tensor
are not degenerate for every atom in the system, then the local coordinate
system $\{\hat{e}_{i}\}$ for each atom, can be uniquely written in terms of
the external global coordinate system $\{\hat{X}_{1},\;\cdots ,\;\hat{X}%
_{d}\}$ that we choose to project $J_{ij}$ on. Calculation of the fraction $%
f $ (in $\%$) of pairs of eigenvalues with $|t_{i}^{\alpha }-t_{i}^{\beta
}|<\delta $ where $\alpha \neq \beta \;\;(\alpha ,\beta =1,\dots ,d)$,
yields that for $\delta =0.5$,$\ \ f\approx 1\%,\;0.2\%\,$and$\;2-3\%$ for
3DbLJ with $\rho =1.2$, 3DbLJ with $\rho =1.4$ and 2DbLJ with $\rho =0.596,$
respectively. For 2DbLJ with $\rho =0.735$, $\ f<0.1\%$. Taking $\delta =0.1$
yields $f<0.1\%$ for all systems except 2DbLJ with $\rho =0.596$ for which $%
f\approx 0.1-0.15\%$. These calculations indicate that the degeneracy
increases when the density decreases. Nevertheless, the actual values of $f$
are very small for all the systems considered, proving that the probability
of degenerate pairs of eigenvalues is negligible.

Next, we compare between the distribution of the $N$ sets $\{t_{i}\}$ (for
each atom $i$ out of the $N$ atoms in the sample) and the distribution of
the $N$ sets $\{t_{r}\}$, where the $d$ numbers $t_{r}^{1},\cdots ,t_{r}^{d}$
are randomly chosen from the measured distribution $P(\left\{ t_{i}\right\}
) $. For each set of $d$ numbers, we choose two numbers according to $%
t_{\min }=\min (t^{1},\cdots, t^{d}),\quad t_{\max }=\max (t^{1},\cdots, 
t^{d}),$ and $%
\left( t_{r}\right) _{\min }=\min (t_{r}^{1},\cdots ,t_{r}^{d}),\quad \left(
t_{r}\right) _{\max }=\max (t_{r}^{1},\cdots ,t_{r}^{d}),$ respectively\cite%
{remark}. Contour plots of the non-normalized distributions $%
P(\{t_{min},t_{max}\})$ and $P(\{\left( t_{r}\right) _{\min },\left(
t_{r}\right) _{\max }\})$) for the systems considered (all at zero
temperature) are shown in figures 1--4. The solid lines correspond to $%
t_{min}=t_{max}$. For the 3DbLJ glasses, each contour plot was calculated
using one sample with $N=16384$ for every system. Similar contour plots were
obtained for other samples of the system (at the same density). In order to
improve the resolution, in the case of 2DbLJ glass we considered sets with $%
N=16384$ and combined four samples for each system (we verified that $%
P(\{t_{1},\;t_{2}\})$ are similar in these sets). Comparing the contour
plots of 3DbLJ $\rho =1.2$ and $\rho =1.4,$ and 2DbLJ $\rho =0.596$ and $%
\rho =0.735$ systems (figures 1.a, 2.a, 3.a and 4.a, respectively), with the
contour plots of the corresponding random sets (figures 1.b, 2.b, 3.b, 4.b,
respectively), we find that the distribution of eigenvalues of the ABT
tensor differs from that of the random sets. In particular, the
non-degeneracy of eigenvalues $\{t_{min},t_{max}\}$ is very pronounced in
the simulated 3DbLJ and dense 2DbLJ glasses, but not in the corresponding
random sets. We conclude that the observed non-degeneracy of eigenvalues
belonging to the same atom is the consequence of a physical constraint, and
not just a trivial consequence of the small probability to randomly choose $%
d $ numbers with close values.

Examining the simulated and random histograms, one observes that the peak of 
$P(\{t_{min},t_{max}\})$ is considerably higher than the peak of $P(\{\left(
t_{r}\right) _{\min },\left( t_{r}\right) _{\max }\})$. The differences
between the most probable ''simulated'' and ''random'' combinations of $%
\{t_{min},t_{max}\}$, can be estimated by comparing the position and the
height of ''random'' and ''simulated'' peaks. Those values for the systems
in figures 1-4 are summarized in the table below. The fraction $f$ of atoms
with $P(t_{min},t_{max})>2/3h_{s}$, where $h_{s}$ is the height of the
simulated peak, is also presented: 
\begin{equation*}
\begin{tabular}{l|c|c|c|c|c|c|}
\hline
& \multicolumn{2}{|c|}{\textit{peak coordinates}} & \multicolumn{2}{c|}{%
\textit{peak height}} & \multicolumn{2}{c|}{$f(P>2/3h_{s})$} \\ \hline\hline
& simulated & random & simulated & random & simulated & random \\ \hline
1 & (-9, 5) & (-7, 4) & 100 & 80 & 0.27 & 0.06 \\ \hline
2 & (-50, -12) & (-44, -14) & 75 & 50 & 0.24 & 0.003! \\ \hline
3 & (1, 3) & (1, 1.5) & 200 & 190 & 0.36 & 0.19 \\ \hline
4 & (-44, -32) & (-42, -40) & 100 & 90 & 0.20 & 0.13 \\ \hline
\end{tabular}%
\end{equation*}%
Comparison between the peak coordinates indicates that the most probable $%
\{t_{min},\;t_{max}\}$ are similar for the simulated and the
corresponding random distributions. For the 2DbLJ and 3DbLJ with the
lower densities, the corresponding $t_{min}$ ($t_{max}$) have small negative
(positive) values, indicating that the average atomic bond tensions are
small. For the dense systems, the corresponding values of $t_{min}$ and $%
t_{max}$ are both negative, reflecting the compression of typical atomic
environments in these systems. On the other hand, comparison of the peak
heights and of the fraction of atoms in the vicinity of peaks, demonstrates
that the probability for the preferred combinations of $\{t_{min},t_{max}\}$
that was measured for simulated systems, can not be reconstructed by a
random choice of eigenvalues.

There are also marked differences in the shapes of the distributions: the
contours of $P(\{t_{min},t_{max}\})$ resemble a cone that expands towards
smaller values of $t_{min}$, whereas those of $P(\{\left( t_{r}\right) _{\min
},\left( t_{r}\right) _{\max }\})$ tend to have a more circular form. To
quantify the difference between the respective histograms, we calculated the
Pearson product moment correlations 
\begin{equation*}
\rho _{X,Y}=\frac{\sum_{i}(X_{i}-\overline{X})(Y_{i}-\overline{Y})/n}{\sqrt{%
\sum_{i}(X_{i}-\overline{X})^{2}/n}\sqrt{\sum_{i}(Y_{i}-\overline{Y})^{2}/n}}
\end{equation*}%
between $t_{min}$ and $t_{max}$ and between $\left( t_{r}\right) _{\min }$
and $\left( t_{r}\right) _{\max }$. The obtained values are (with last
significant digit $0.05$): 
\begin{equation*}
\begin{tabular}{|l|c|c|}
\hline
$\;\;\;\;\;\;\;\;$ \textit{system} & $\rho (t_{min},t_{max})$ & $\rho
(t_{min}^{r},t_{max}^{r})$ \\ \hline\hline
3DbLJ ($\rho =1.2$ ) & $0.7$ & $0.3$ \\ \hline
3DbLJ ($\rho =1.4$ ) & $0.75$ & $0.3$ \\ \hline
2DbLJ ($\rho =0.596$) & $0.75$ & $0.5$ \\ \hline
2DbLJ ($\rho =0.735$) & $0.7$ & $0.5$ \\ \hline
\end{tabular}%
\ \ 
\end{equation*}%
Although the shapes of the distributions (compare figures 1a and 2a for the
3d case, and figures 3a and 4a for the 2d case) and the correlations between 
$t_{min}$ and $t_{max}$ are quite similar for the densities studied, the
dispersion in the values of $t_{min}$ and $t_{max}$ is much larger for the
denser systems (notice the different scaling of the axes in the above
figures).

The observed high correlations show that if $t_{min}$ has a small (negative)
value then it is highly probable that $t_{max}$ is also small. Vice versa,
if $t_{min}$ is large we expect $t_{max}$ to be large too. The absence of
local environments with $t_{min}\ll 0$ and $t_{max}\gg 0$ in the simulated
systems, suggests that \emph{the local organization of the eigenvalues of
ABT tensor is more isotropic than if they were randomly distributed. }On the
other hand, the previously discussed observation that the eigenvalues
belonging to the same atom are always non-degenerate, implies that this
tendency towards local isotropy of atomic bond tension is frustrated, and
that the local stresses in the glass state can not be completely
characterized by isotropic pressure.

\section{Directional correlations and clustering}

Having shown that the eigenvalues of ABT tensor belonging to the same atom
are correlated on average, we address a related question of whether atoms
with similar eigenvalues $t_{i}$ are clustered in space. Since for most of
the atoms the tensions in their bonds are quite small, the issue of
clustering is mostly relevant for atoms with extreme (high and low) atomic
bond tensions, i.e. for the tails of the distribution of $P (t_{i}) $. Using
different samples of 2DbLJ and 3DbLJ glasses at zero temperature, we looked
(for each sample) for clusters of atoms with $t_{min}$ ($t_{max}$ ) below
(above) some given threshold. We did not observe clustering of atoms with
high $t_{max}$ in any of the samples. For dense 2DbLJ glass and for all
3DbLJ glasses studied, atoms with the lowest (negative) values of $t_{min}$
were found to cluster together.

A simulated realization of a dense 2DbLJ glass in mechanical equilibrium is
presented in figure 5 (both types of atoms are drawn with the same size).
The  $33\%$  of atoms with the lowest (negative) values of $t_{min}$ are marked 
with yellow color and the remaining $67\%$ are colored in blue, light blue, white, pink and red,
 corresponding to increasing values of $t_{min}$. As can be seen from the figure, the yellow
atoms form nearly one dimensional percolating chains. Similar 
pictures were obtained for other samples. The measured percolation threshold for 
atoms
with lowest $t_{min}$ is $\approx 30\%$. This value is lower than the
theoretical percolation threshold for two dimensional lattices\cite{Stauffer}. 
Indeed, for 
randomly chosen $30\%$ of atoms, we found considerably smaller clusters (with 
less atoms), 
that did not percolate.

Repeating the above measurements (with different thresholds) for snapshots
of dense 2DbLJ glasses at higher temperatures, we did not observe such
''force chains''. On the other hand, running these systems at temperatures
well within the vibrational regime (i.e. when no hopping occurred), and then
performing a steepest descent quench to freeze the atomic positions at local
energy minima (see details in \cite{thesis}), we observed similar patterns.
This observation suggests that the mechanical equilibrium condition of zero
force on each atom is a necessary condition for the formation of percolating
force chains.

To further investigate this point, a different preparation scheme was used
to generate dense 2DbLJ systems. Using samples of frozen 2DbLJ glass with $%
\rho =0.596$, all atomic positions and the length of the simulation box were
rescaled by $10\%$, then the system was frozen again to its new local
minima. Comparing between the initial (with $\rho =0.596$) and final (with $%
\rho =0.735$) samples, we observed that atoms with smallest $t_{min}$ are
not the same in the original and the final states. We also observed that
such compression alone can not account for the formation of force chains.
Only after relaxation to a state of local mechanical equilibrium, the
resultant sample formed ``force chains'' similar to those previously
described.

A percolating cluster of atoms with lowest values of $t_{min}$ in 3DbLJ
glass at zero temperature is shown in figure 6 (to improve visualization,
only a single cluster is plotted). If the threshold for $t_{min}$ is chosen
to include $\approx 5-6\%$ of atoms (we found that with such a threshold the
system is near to or slightly above percolation), the percolating clusters
form nearly one dimensional chains and loops. Although the threshold values
of $t_{min}$ are different for the compressed and non-compressed 3DbLJ
glasses ($t_{min}<-77.5$ and $t_{min}<-23$, respectively), we did not
observe qualitative differences in the force chains.

From the observations described in the previous paragraphs, we conclude
that both the mechanical equilibrium condition of zero net force on every
atom and small values of $t_{min}$, are necessary for the formation of force
chains. To see why this is so, recall that mechanical equilibrium implies
that the components of the force on an atom along mutually orthogonal
directions have to vanish. In particular, for $\alpha =1,\dots ,d$ 
\begin{equation}
f_{i}^{\alpha }=\sum_{j=1}^{N}J_{ij}\hat{r}_{ij}\cdot \hat{e}_{i}^{\alpha
}=0,  \label{forcecons}
\end{equation}%
where $f^{\alpha }$ is the component of the force acting along the
eigenvector $\hat{e}^{\alpha }$. The environment of a typical atom $\ i$
(corresponding to the vicinity of the peak of the distribution shown, e.g.,
in figure 1a) consists of mildly compressed and mildly stretched bonds, and
is therefore nearly isotropic. In this case, the mechanical equilibrium
condition, equation \ref{forcecons}, can be satisfied by balancing the
negative and positive tensions in the bonds of the atom with all its
neighbors. Now, consider an atom with 
\begin{equation*}
(t_{i})_{\min }=\min \sum_{j=1}^{N}J_{ij}\left( \hat{r}_{ij}\cdot \hat{e}%
_{i}^{\alpha }\right) ^{2}\ll 0.
\end{equation*}%
that belongs to the tail of the distribution.
Inspection of figure 1a reveals that for $t_{\min }\ll 0,$ the spread of
values taken by $t_{\max }$ (and, in 3d, by the intermediate eigenvalue)
increases with decreasing $t_{\min }$ resulting in increasingly anisotropic
atomic environment. In this case, the condition of zero force on atom $i$
can only be satisfied by two similarly compressed bonds with atoms $j$ and $%
k,$ oriented at nearly opposite directions. In order to balance the forces
on atom $j$ there should be another nearest-neighbor with a highly
compressed bond in a nearly opposite direction, and so on, leading to the
formation of force chains. Note that above arguments imply that force chains
would also be observed if one considers extremal normal forces $J_{ij}$
instead of extremal eigenvalues of the ABT tensor $t_{i}$. This result is in
accord with that of ref. \cite{Wittmer}.

The above arguments can also explain, why for two and three dimensional
binary LJ glasses, similar formation of force chains of atoms with high
values of $t_{max}$ is not likely (although in ref. \cite{Wittmer} weaker
force chains were observed for pairs of highly stretched bonds). For these
systems, the eigenvalues of ABT tensor corresponding to a highly stretched
atomic environment are of similar magnitude (see figures 1a, 2a, 3a and 4a)
and force  balance is achieved by nearly isotropic stretching of bonds
between nearest neighbor atoms. Such a nearly isotropic organization implies
that even if atoms with high $t_{max}$ were to cluster, the clusters would
not take the form of nearly one-dimensional chains, but rather that of
isotropic localized inclusions. These results demonstrate that the
directional organization of highly stretched atomic environments is
different from that of highly compressed ones. Since $t^{\alpha }$ are
related to the internal stresses that are present in glassy systems even in
mechanical equilibrium, our numerical results agree with Alexander's
conjecture about the different organization of positive and negative
internal stresses in glasses (the atoms with high values of $t_{max}$
correspond to the ''soft inclusions'' discussed in ref. \cite{Alexander}).

\section{Discussion and conclusions}

In this paper we defined a new parameter, the atomic bond tension (ABT)
tensor, and used it to study the directional organization of atomic bond
tensions. We found that in sufficiently dense systems at very low
temperature, the eigenvalues $\{t_{1},\cdots ,t_{d}\}$ of ABT tensor are
correlated i.e. are on the average, similar in sign and magnitude. Thus, if
one observes that the environment of an atom is stretched (compressed) along
some direction, one can predict with high probability that it would be
compressed (stretched) along other directions as well. Based on this
correlation, one may conclude that atomic environments in glasses in
mechanical equilibrium are isotropic, at least in a statistical sense.
However, since the eigenvalues of the ABT\ tensor of an atom are strictly
non-degenerate, we conclude that the tendency towards perfect isotropy of
atomic environments is frustrated.

Since for systems with radial pair potential there is one-to-one
correspondence between the separation $R_{ij}$ and the tension $J_{ij}$ in
the bond $[ij]$, the observed correlations between the bond tensions 
imply also correlations between the corresponding $R_{ij}$. Thus,
various combinations of inter-atomic distances, and especially of
nearest-neighbor separations in glassy systems, are constrained, and their
local organization is not random. The observed deviations from randomness
are associated with directional correlations; if only isotropic measures of
disorder such as local volume strain \cite{Egami97},  deviation from local
sphericity \cite{Starr} or local pressure \cite{Kustanovich2} are
considered, such correlations are not taken into account and a more random
description of dense liquids and glasses results. 

The results of numerical simulations reported in this paper provide a direct
confirmation of Alexander's conjecture that ``the reference state of the
glass cannot be regarded as an arbitrary random state'' (see chapter 16 in
ref. \cite{Alexander}). Qualitatively,
correlations among ABT are expected from the requirement that the forces on
each atom of a structurally disordered system in mechanical equilibrium are
balanced, and that each atom is caged in a potential well, i.e., that each
atom is subject to a set of constraints that include zero net force and a
stable atomic environment. For systems with a radial pair potential, the
one-to-one correspondence between the separation of two atoms and the
tension on their ``bond'' means that achieving balance of forces and local
stability simultaneously for every atom in the system, reduces the number of
degrees of freedom and sets strong constraints on the observed ABT. The
observed strong correlations between the eigenvalues of ABT, in two- and
three-dimensional binary LJ glasses, imply that when the asymmetry between
the bonds is strong (e.g., due to two types of atoms), the directional
organization of bond tensions that would comply with force balance and local
stability is subject to stronger constraints.

Since the Einstein frequencies of atomic vibration around an equilibrium
position are determined by the derivatives of the inter-particle pair
potential, the observation of correlations between the corresponding $R_{ij}$
implies that the atomic force constants are not random as well. Thus,
further quantification of correlations in atomic force constants is required
before one can incorporate them into lattice-based models which aim to
reproduce the inhomogeneous network of force constants  measured for simulated
glasses (see, e.g., \cite{Angelani} ).

The self-organization of normal forces under various external stresses
(compression, shear) has been extensively studied for granular materials %
\cite{Cates,Tkachenko,Radjai99,Bouchaud}, and recently
also for liquids and glasses \cite{OHern},  and for model
random packing \cite{OHern1}. We studied the self organization of bond
tensions on an atomic level, and showed that external compression has strong
qualitative effect on the local organization of the atomic bond tensions. In
compressed two and three dimensional binary LJ glasses, atoms with the
smallest eigenvalues of ABT tensor are organized in nearly one dimensional
chains and loops. Since force chains were observed only for systems at zero
temperature, we concluded that balance of forces on each atom is a necessary
requirement for such organization. Although this is a customary requirement
in modeling of force chains in granular materials \cite{Liu}, to the best of
our knowledge such a requirement, although obvious, was not previously
suggested in the context of glasses. In glasses with repulsive--attractive
pair interactions, force balance can be achieved both by directional
organization of stretched and compressed bonds in a way that their total
effect is zero, or by strong, nearly collinear compression of pairs of bonds
(similar stretching of pairs of bonds is not possible, since it is not a
stable configuration). For externally compressed glassy systems, there are
not enough stretched bonds to balance the  compressed ones, and only
self-organization into one-dimensional structures (force chains) of highly
compressed bonds can balance the forces for every atom.

Finally we would like to comment on several open issues for future research.
We do not yet understand the physical origin of\ the observed conflict
between the directional correlations and the strict non-degeneracy of the
eigenvalues of ABT, and of the resulting frustrated tendency towards local
isotropy of atomic environments in glasses. While we demonstrated the
existence of organization of the atomic bond tensions in glasses and in
Inherent States of liquids and quantified the extent to which they deviate
from a random distribution, a more intuitive understanding of the difference
between such complex organized states and purely random ones, is clearly
needed.
\subsection*{Acknowledgments}

YR would like to acknowledge the support by a grant from the Israel Science
Foundation. \\
\\
Zeev Olami died before this manuscript was submitted. We miss him a lot.

\section*{Figures}
\begin{enumerate}
 \item 
a. Histogram of $P(t_{min}, t_{max})$ for 3DbLJ system with $\rho=1.2$ at $T=0$ 
($N=16384$). b. Histogram $P(t^r_{min}, t^r_{max}) $ for the same 3DbLJ system.

\item 
a. Histogram of $P(t_{min}, t_{max})$ for dense 3DbLJ  system with 
$\rho=1.4$ at $T=0 $ ($N=16384$ particles). b. Histogram $P(t^r_{min}, t^r_{max})$
 for the same dense 3DbLJ system.

\item 
a. Histogram of $P(t_{min}, t_{max})$ for 2DbLJ systems with $\rho=0.596$ 
at $T=0$. (Total number of particles is 16384.) b.
Histogram $P(t^r_{min}, t^r_{max})$ for the same 2DbLJ systems.

\item 
a. Histogram of $P(t_{min}, t_{max})$ for dense 2DbLJ systems with $\rho=0.735$ at 
$T=0$. (Total number of particles is 16384.) b. Histogram $P(t^r_{min}, t^r_{max})$ 
for the same dense 2DbLJ system.

\item 
$t_{min}$ for a sample of dense 2DbLJ system at zero temperature.
Atoms marked in yellow are those with $t_{min} \le -56$ (this subset
contains $33 \%$ of atoms with lowest $t_{min}$).

\item 
3DbLJ systems with $\rho=1.4$ - loop-like cluster of atoms
with $t_{min}<-77.5$.
\end{enumerate}


\begin{thebibliography}{99}
\bibitem{remark} In nearly isotropic atomic environment, the eigenvalues $%
\{t^{1},t^{2},t^{3}\}$ have close values and considering only the extreme
values $t_{min}$ and $t_{max}$ captures this point. If the atomic
environment is highly compressed in one direction (corresponding to $t_{min}$%
) and highly stretched in an orthogonal direction (corresponding to $t_{max}$%
), there is a considerable range of values that the third eigenvalue can
take, and important information may be lost by considering only $%
\{t_{min},t_{max}\}$. For the 3DbLJ systems considered, both simulated and
random distributions, the probability for having simultaneously $t_{min}\ll
0 $ and $t_{max}\gg 0$ was found to be vanishingly small, thus validating
our approach.
\end{thebibliography}
\end{document}